# The Structural Dynamics of Corruption
## Artificial Society Approach

**Hokky Situngkir**
(hokky@elka.ee.itb.ac.id, quicchote@yahoo.com)
Department of Computational Sociology
Bandung Fe Institute

## Abstract

Corruption has been an important issue as it becomes obstacle to achieve the better and more efficient economic governmental system. The paper defines corruption in two ways, as state capture and administrative corruption to grasp the quintessence of the corruption cases modeled in dynamical computational social system. The result of experiments through simulation is provided in order to construct an understanding of structural properties of corruption, giving way to consider corruption not as an isolated phenomenon, but conclusively, as an interdisciplinary problem and should be handled in holistic perspectives.

> *….Now kings will rule and the poor will toil*
> *And tear their hands as they tear the soil*
> *But a day will come in this dawning age*
> *When an honest man sees an honest wage…*
> Van Diemen's Land – U2

## 1. Corruption as Floated Signifier

It is obvious that corruption is not an exclusively economic phenomenon. As described by Abed & Davoodi (2000), corruption manifests itself in many political processes, law and judicial system, and many less visible spheres. However, the most popular and probably simplest definition was coming from Tanzi (1998) that defined corruption as "*the abuse of public power for private benefit*".

Corruption, however, is a difficult term to define. Gambetta (2000):

> Can we identify a specific social practice that we can justifiably call 'corruption', and, if so, what are its distinct analytical properties? Given the multiplicity of definitions found in the literature and the considerable confusion over what exactly we should understand corruption to mean, this question, which forms the object of this essay, does neither have a straightforward or a formalistic answer.

In many cases, corruption is a signifier referring to many phenomena. Most of literatures on corruption refer to many phenomena i.e.: bribery, collusion, nepotism, and so on. According to Berg (2001), corruption is the abuse of public power for private benefit, while the private benefit is often in the form of illicit money or in-kind from a client to the agent; we call this as bribery. Conclusively, the evaluation of the definition on corruption from many literatures evokes us to realize that the terminology of corruption is a floated signifier many terminologies should have point out.

In advance, we can say that corruption is a symptom of the weakness of political, social, legal and economic systems. Even where corruption is widespread, the actor will strive to keep it hidden from public view. Corruption is not new, nor is it confined to any particular part of the world. On the contrary, corruption is a global phenomenon, although its severity varies from country to country.

One important thing to note is that the different definitions on corruption eventually impact the way we measure and analyze a corrupt phenomena. Berg (2001) analyzed some different methods on measurement on corruption. Today, corruption is measured through surveys and polls of random sample from citizens or businesspeople; this is what can be called the method of subjective measures, although each of the subjective measurement employ distinguishable methodologies: the first based on the perceptions and the other based on experience. However, the paper will not discuss the methodologies of the measurement since the purpose of the paper is to find some general structural outline of corruption.

The rest of the paper will construct analytical tools from the definition of corruption directed toward some practical cases. We will use analytical map of corruption (Gambetta, 2000) and the game-theoretic model of corruption in bureaucracies (Norris, 2000). Eventually, we will try to construct the dynamical model based on the definitions and the classifications described in the next section.

## 3. What can we call corruption?

To define what corruption is, we must analyze the social agent that involved in the process. According to Gambetta (2000), there are at least three agents involved in the corruption: agent (can be individual, such as an employer, or a collective body) relying on the expectation that people in certain positions are bound to follow given rules, one who agrees to act on behalf of the first agent, and the other whose interests are affected by the second agent's actions. There are rules the second agent must obey as trusted by the first agent. However, the third agent wants the second agent to do improperly: to abuse the second agent's trust from the first in order to gain benefit for her. For the sake of easier linguistic environment we will give three set of agents, the first agent is member of the set *T*, to give trust to the second agent. The second agent is member of the set *G* receiving obligation to play according to rules that tie *G* and *T*. However, during the process there will be the third agent, the member of the set *P*, try to attract member of *G* to violate the rule specified. Once member of *G* agrees to do it, the corruption on abusing the authority occurs. We should note that a corruption can only occur if and only if a certain relationship between members of *T* and members of *G* pre-exists.

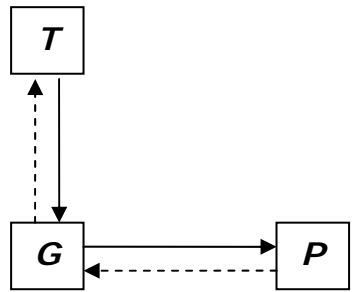

**Figure 1**
The primitive model of corruption.
The solid arrows represent things to be done according to rule of the game and the other represent situation when corruption occurs.

In our case members of *T* agree to entrust some resources or objects to members of *G* if the members of *T* expects the members of *G* to serve their interests. The members of *T* trust the members of *G* to do things whose outcome is *x* for *T*. The wrong deeds abusing the trust by members of *G* denoted by *y* will reduce outcome the members of *T* should obtain, since *y*>*x*. Obviously by abusing the trust, members of *G* gain more benefit. An act of corruption is then defined by the amount of some currency, say *b\**, received by members of *G* being persuaded by the members of *P* so the members of *G* abuse the trust they received. We can say that the value of *b\** fulfills

$$y + b^* > x$$

This is the value motivating members of *G* to abuse the trust. Members of *P* can vary the value of *b\** so in return the members of *G* can reject the offer because it is too small relative to the cost they incur by doing the rule, or probably because their integrity: a moral keep them not to abuse the trust. It is important to note that there should be an exchange between members of *P* and *G* then the corrupt action occurs.

Following the notions above, we can say that bribe (use of rewards to pervert the judgment of a person in a position of trust), nepotism (patronage by reason of certain types of primordialism rather than capability), and misappropriation (illegal appropriation of public resources for private regarding use) are aspects of corruption. A bribe to the police officer, the judge, or else can be classified as corruption since the members of *P* offer some currency to the members of *G* to break the law of the supreme power (the government, the representatives of the whole people, etc.). However, we should consider also that sometimes members of *P* offer bribe to *G* in order to have their own rights as citizens. For example, we should offer bribe to a trash-collector so he will take away our trash, or otherwise, he will not serve us well. In some developing countries such case is common.

Such works like Hellman, et. al. (2000), Berg (2001), up to the works of the Transparency International (2003) and The Gallup Institute (1999) practically give us the structured perspectives of common people or citizens on corruption in various countries. The method applied is statistics to measure how the citizens perceived corruption in their society.

One example is the cases of corruption in Indonesia. Researches by the Partnership for Governance Reform in Indonesia (2002) showed that corruption can be classified in two terms:

- ✓ *the Administrative Corruption*, defined as intentional imposition by state or non-state actors to distort existing laws, policies, regulations for their own advantages. In our basic model, this case can be drawn as in figure 2a. There are some rights the citizens should enjoy as guaranteed by the authorities, but the officials abuse their authorities to gain as much advantages as they can.
- ✓ *the state capture*, defined as illegal actions by firms or individuals to influence the formulation of laws, policies, regulations for their own advantage. In many post-communist countries this kind of corruption can differ in two other classifiers (Hellman, et.al., 2000), i.e.: *state capture* (influence with illicit payment) and the *influence* (without illicit or non-transparent payment). In Indonesia, there is no complete literatures alike to justify the distinction the classifier indicates, thus we will use the term *state capture* applied to the rule of the game.

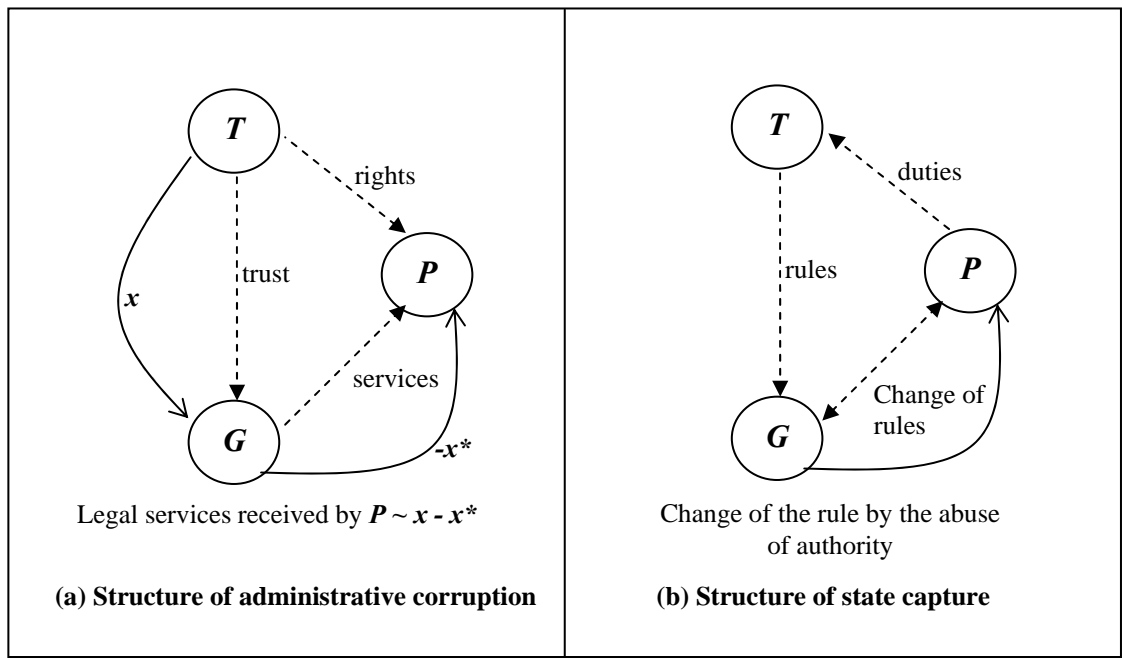

**Figure 2**
The general structures of corruption occur in Indonesian society

Seeing the figure 2, we find that corruption in many cases should be viewed through the perspectives of the interactions between the members of **G** and **P**: there will always be two sides exchanging benefit and abusing the trust and authority pre-gained by one of them. In cases of administration corruption we can see bribe as the generic model. It affects someone's right, the members of **P** whose connection with **T** must be actualized by interacting with **G**. In this type, the members of **G** abuse their positions and authority to gain benefit from the offer of the members of **P**. As explained above, the offer can be actualized for breaking the rule of service of members of **P** by members of **G** so it reduces the right to have service of members of **P**.

The second type is mostly found in unclear political system where members of **P** can influence the rule of the game contracted by the members of **G** and the members of **T**. The cases usually happen on the duties that should be fulfilled by members of **P** for the members of **T**, but in order to accomplish the duties, members of **P** should interact with the members of **G**. By actualizing this mode of corruption, the members of **P** will reduce the duties they should accomplish according to the rule since they will influence the members of **G** to use their authority on changing the rule to give the most benefit they can get.

Somehow, we can find the terminologies on collusion and nepotism inherently in the two types of corruption. It is not an exaggeration to say that they are the generic models of the practical corruptions. In the next section, we approach the dynamics and evolutionary corruptions by the two generic types. The static game-theoretic model of bureaucratic corruptions (Norris, 2000) can be seen as the more technical views of the two types of corruptions.

## 4. The Dynamics of Corruption: Model Construction

From the two types of corruption explained in the previous sections we understand that corruption can be seen as general form described in figure 1. Practically, we can classify corruptions as seen in figure 2; how the abuse of power misappropriates the rights or the duties contracted by the rule of the game. We construct a dynamical model to see some transitions or some changes in the social system regarding the corruption cases within. We build model based on the general model described on the figure 1 on varying parameters of social system by realizing that the figure 2 will only give us some details that can be seen succinctly in the general model.

There are three groups of agents we will deal with. We isolate two interacting sets of constituents in the model (members of **G** and **P**) as corruption rises from these two sets of agents. The two set of agents interact each other in random order and sequences while there will be exchange of benefit between the two of them corrupting the rule of the game. We build subjectivity in every agent, i.e.: the memory to remember what has happened with her friends in the previous iterations, the preferences based on her morality, subjectivity on the system, whether or not she asks her partner to corrupt or do honestly.

We use the simple pay-off matrix presented by Hammond (2000) with some modifications. In the model, we construct two sets of agents, representing members of **G** and the members of **P**. Each agent from each set interacts in the trust game based on the pay-off matrix pre-defined. Their neighbors and partners with whom she interact with is selected randomly as in figure 3. Each will get pay-off as the consequence of the choice or decision. Corruption will only occur when the two partners agree to do so. We should note that the agents are bounded rational on their expectations of the next turn choice.

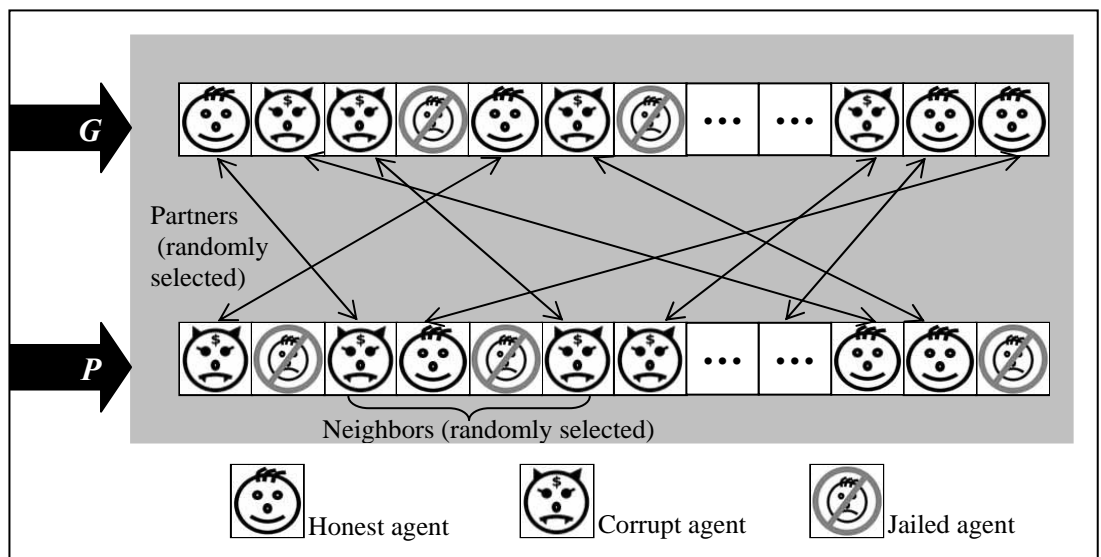

**Figure 3**
The way the game roles.

There are two subjective considerations on every agent to choose whether to corrupt or not, i.e.:
- subjective thought whether or not her partner in the next round will agree to corrupt or not, denoted by **F**, based on the number of the matched corrupt partner that agreed to corrupt (e.g.: to accept or offer bribe) per the length of memory of each.

$$F = \frac{\text{Matched corrupt Partners}}{\text{Memory}}$$

- subjective thought whether or not she will be caught by the members of **T** that can sentence her to the jail for some certain rounds, **C**, assumed by the jailed friends (**m**), in her social network per corrupt friends in the last round (**M**).

$$C = \frac{\text{Friends in Jail}}{\text{Corrupt Friends}}$$

As figured in table 1, the biggest payoff value will be gained if both players choose to corrupt. Every agent in our system will be pre-given the honesty index, the index shows how corrupt one to be. Practically, we give the index by [0, 1] randomly among agents; the more dishonest an agent to be corrupt the closer the index to unity.

Table 1
The Pay-off Matrix of the Game

|  | Corrupt | Not-Corrupt |
|---|---|---|
| **Corrupt** | $\alpha$ | $\beta$ |
| **Not-Corrupt** | $\beta$ | $\beta$ |

Honesty index determines the expectation of each agent on every round's pay-off. The closer the index to unity, the bigger the expectation of the agent, or the greedier the agent as follows:

$$\alpha^* = (1-i)\alpha$$

where the **α*I** is the expectation of the corruption and **i** as the honesty index.

By glancing at the pay-off matrix we can see that the game seems to be very simple, since **α > β**. By assuming the game as static, we can solve it and find the equilibrium, but under dynamical system, the game will show the complexity of the game, since there will be emergence in the collective state of all agents. In other words, the complexity of the game comes out from the uncertainty of each agent to choose whether corrupt or not.

Eventually, the agent uses limited information (the agent never knows and calculates the macro-state of the system) to choose as follows:

$$E(x) = (1-C)\left|F\alpha^* + (1-F)\beta\right| + C\left|\beta - k\beta\right|$$

where **k** is the length of the jail term. If the expectation **E(x) > β** then the agent is greedy enough to do corruption. It is obvious that the agent builds up subjective perspectives on her environment to choose to corrupt or to be honest. The simulations of this model will bring us to the ability to answer some general points on corruption cases in society analytically.

# 5. Results of Simulations

We do several experiments on the artificial corrupt society. The list of basic numerical variables is available in the appendix of the paper. We do some experiments based on the structures and some structural alternative solutions oftenly proposed to handle corruption. Our first experiment wants to test how the social system gets its evolutionary stable in which the variables used are at the basic values.

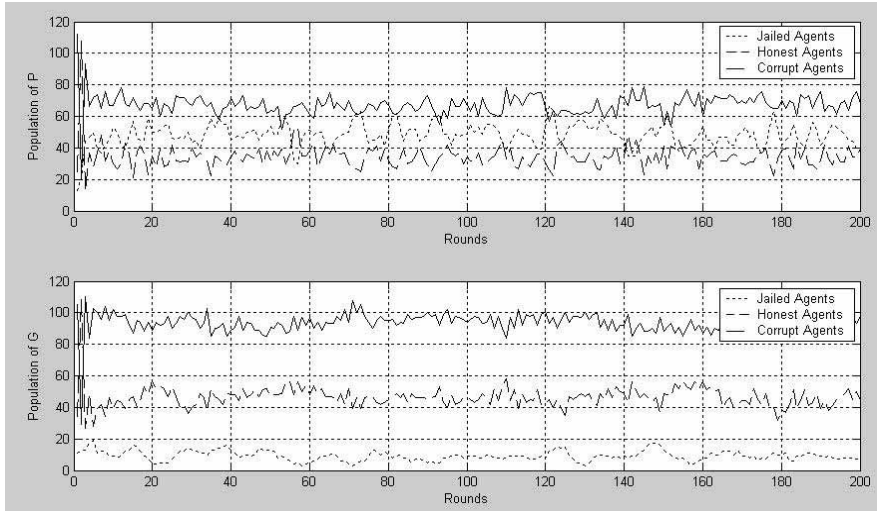

**Figure 4**
The first simulation result using the basic parameters on the artificial society

In figure 4, we can see that the population of ***P*** jailed more than the members of ***G***, since in most cases of corruption, ordinary people is much easier to get caught of doing corruption than those who are members of the bureaucrats. The evolutionary stable conditions seem to make the corruptors remain to corrupt while the honest ones remain honest or are stimulated to corrupt. The simulation result reflects the algorithm employed here that the jailed corruptors will perform honestly right after she leave the jail while she can be corruptor again in the next round. That is why the population of corruptor is seemingly more stable than others in both graphs.

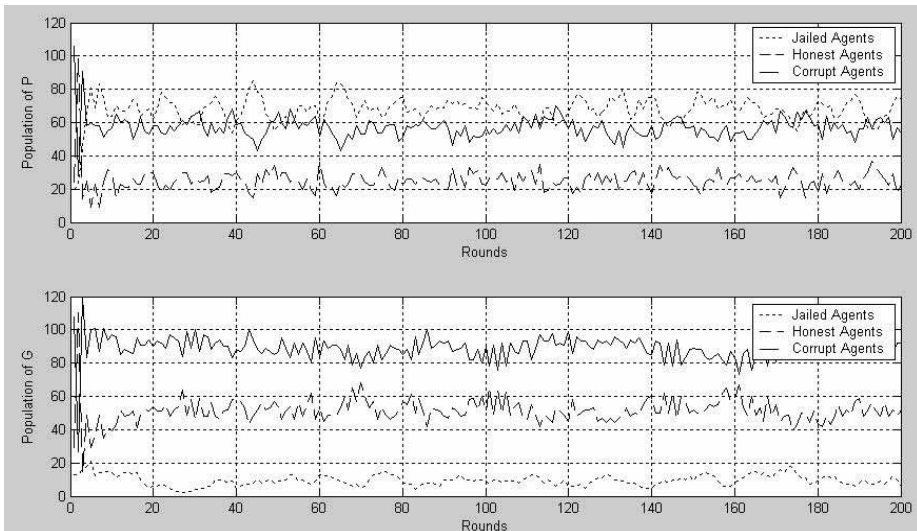

**Figure 5**
The performance of the system on the same jail period. There is a fast transition on the domination of honest agents and the corrupt one in the populations since the populations of ***P*** learn and fear of corruption the many of their friends jailed on corruption relative to the bureaucrats.

When we give the members of *G* longer length of jail period than the members of *P* as described in figure 6, the members of *G* then constitutes very corrupt bureaucracies with minimum jailed agents. This fact is followed by the dominant corrupt agents among the population of the jailed agents of *P*. This result concludes a common propositions saying that corrupt government will induced corrupt citizens. We should note also, that as has been proved by the previous work (Situngkir, 2003a), agent's morality cannot become a major solution to combat corruption. In our simulation experiments we do not change and analyze the honesty index. As introduced before, the honesty index is pre-defined parameter that will never change during the run of simulations.

In our simulations, the corrupt culture in the members of *G* induces the corrupt behavior among the members of *P* do not care how honest the populations of *P*. This suggests that the combat of corruptions somehow should begin first from the law enforcement regarding all the populations of bureaucracies and then the citizens. The endeavor to combat corruption by campaigning to the people without law enforcement (as described in the next experiments) will eventually fail. This is emphasized by the next simulations in figure 7, where the length of the jail period of members of *P* is longer than the members of *G*. In other words, combating corruption should be done holistically.

There is an alternative solution proposed by the common sense: if we give a chance to giant corruptors then spreading corruption among the whole people will follow. Our simulation shows that this is experimentally correct. When the highest authority (represented in our general model as members of the *T*) jails the giant corruptors (represented by the most often corruptors) there will be fast transitions from the corrupt regimes to the dominations of the honest and the lowest level of corrupted agents. This fact shows us that it is important to put the big corruptors in jail so other agents who corrupt will follow by rarely turning out to be honest.

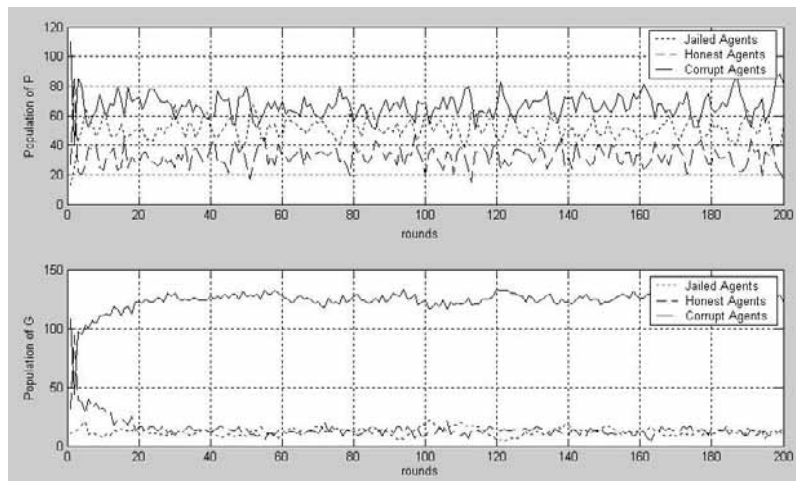

**Figure 6**
The dominations of corruption in members of *G* on the longer jail period than the agents of *P*.
The honest agents of *P* will be induced to be corrupted populations along with the corruptions in the bureaucracies (members of *G*).

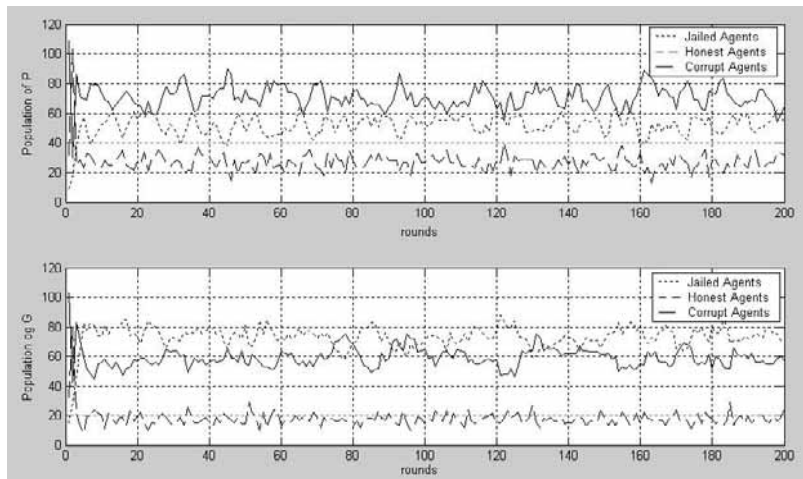

**Figure 7**
The jail period among members of ***P*** is longer than ***G***. Following the previous simulation result, the corrupt citizens are reflected by the dominance of jailed agents among bureaucracies.

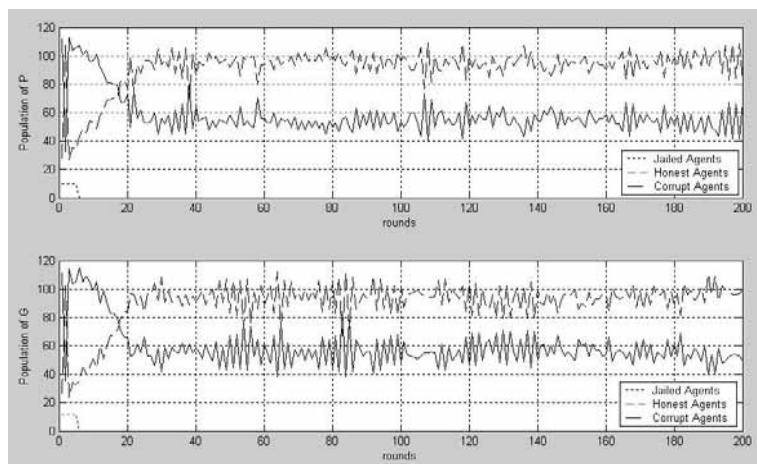

**Figure 8**
The transitions among the two interacting populations. Law enforcement to put the most often corrupting agents in jail shifts the corrupt regime to the honest one.

Eventually, we check computationally in our model whether a corruption regime will fail on the better pay-off for the honest. Practically, this is very popular in some developing countries to say that an important way to reduce corruption is to make the wages higher. However, as described in the beginning of the paper, we must understand that corruption is not merely economic problem. Corruption is linked to many aspects of the life of a society, including political, culture, and sociological aspects. In economics, a comprehensive agent-based computational economics on corruption with the macro-economical aspects can be seen in Chakrabarti (2001). On this paradigm, the solutions to raise the wage comes from the assumption that corruption is merely an economic problem. A nice statistical analysis proving this proposition practically can be seen on the works by Rijkeghem & Weder (1997).

As seen in figure 9, we can see that at raising the pay-off higher for an honest agent relative to the corrupt one will give no solution at all. The corrupt agents will still dominate among the member of ***G*** and ***P***. This fact however demands the more holistic view on corruption and constructing alternatives to combat it. Solving problem of corruption only economically will end up in nothing since corruption has rooted not only in economic properties of social system but also in many other aspects.

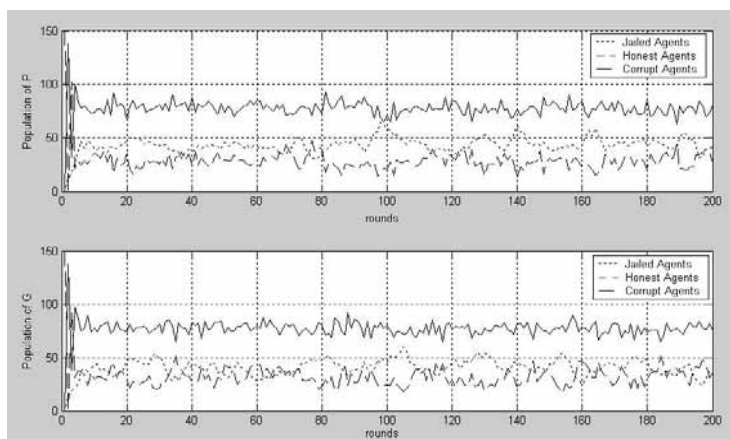

**Figure 9**
The merely higher pay-off (or wage) to do honest relative to the corrupt one is not a good solution to combat corruption. This indicates a need to view corruption more holistically in order to construct the strategy to eliminate it.

# 6. Further Works & Concluding Remarks
The corruption is defined as the abuse of authority and or power trusted to someone by means of gaining self-benefit offered by a third party who finds benefits from the abuse of the trust game. By this analysis we defined the administrative corruption and the state capture as the two basic types of corruption often happen. Then we obtain general model of corruption by seeing the vested interests on abusing the legal proportions on duties and rights of citizens as depicted in both models.

The computational simulation experiments performed shows that the dynamics of corruption touches the two interacting agents, i.e.: the trusted agents and the agents offering the abuse of the trust. Henceforth, the way to cope with corruption should also touch these two agents simultaneously. Forgetting one agent in the basic model will be giving no effect to the elimination of corruption. Thus, to see corruption correctly is to see it as a complex dynamical system in the complexity of social system.

It is understood that the rule-mechanism of each agent is seemingly simple since the model presented here is constructed on the purpose of clarifying the general structural dynamics of corruption. To make the attributes of the agents to be more complicated in order to find more emergent phenomena is a challenging further work to be considered, i.e.: the use of some macroeconomic variables e.g.: inflation, economic growth, etc. This further works will enrich any policies made for reducing corruption practically.

In general, we have shown that corruption should not also be seen in mono-dimensional spectacles. The corruption is not merely economic phenomena but deeply rooted in many aspects of social dynamics, politically and economically embedded inside the culture of the entire people in all level of social hierarchies. Therefore, to eliminate corruption is to construct holistic strategies involving as wide as possible field of discourse. Corruption, however, is an interdisciplinary field of discussions.

## Acknowledgement:

I would like to thank Dodi Rustandi for discussions and motivations in which period the research is done, member of Board of Advisory BFI, Yohanes Surya, for the helpful financial support, and two anonymous referees for their encouragement and constructive suggestions on the draft version of the paper. All faults remain my own.

14. Keefer, Phillip. (2002). *The political economy of corruption in Indonesia*. DECRG. URL: http://www1.worldbank.org/publicsector/anticorrupt/FlagshipCourse2003/KeeferIndonesia.pdf
15. Lindgren, Kristian. (1996). *Evolutionary Dynamics in Game-Theoretic Models*. Working Paper WP96-06-043, Santa Fe Institute.
16. Makarim, N.A. (2001). *A Path through the Rainforest: Anti-Corruption in Indonesia*. In Asia Program Special Report No.100/December 2001. Woodrow Wilson International Center for Scholars. URL: http://www.wilsoncenter.org
17. Marjit, Sugata. & Mukherjee, Arijit. (1996). *A Simple Theory of Harassment and Corruption*. URL: http://netec.mcc.ac.uk/pub/NetEc/RePEc/remo/bon/bonsfa/bonsfa527.ps
18. Myerson, Roger B. (1991). *Game Theory: Analysis of Conflict*. Harvard University Press.
19. Norris, Era-Dabla. (2000). *A Game-Theoretic Analysis of Corruption in Bureaucracies*. Working Paper WP/00/106. International Monetary Fund.
20. Partnership for Governance Reform in Indonesia. (2002). *A Diagnostic Study of Corruption in Indonesia*. Final Report. URL: http://www.partnership.or.id/
21. Rijkeghem, Caroline Van., & Weder, Beatrice. (1997). *Corruption and the Rate of Temptation: Do Low Wages in the Civil Service Cause Corruption?*. IMF Working Paper WP/00/106. International Monetary Fund.
22. Situngkir, Hokky. (2003a). *Moneyscape: A Generic Agent-Based Model of Corruption*. Working Paper WPC2003. Bandung Fe Institute.
23. ______________. (2003b). *NGOs and The Foreign Donations: The Possibilities of Fuzzy Corruption in The Fuzziness of Social Empowerment(?)*. Working Paper WPN2003. Bandung Fe Institute.
24. Situngkir, Hokky. & Hariadi, Yun. (2003). *Dinamika Evolusioner Kontrak Sosial di Indonesia*. Working Paper WPK2003. Bandung Fe Institute.
25. Tanzi, V. (1998). *Corruption Around the World: Causes, Consequences, Scope, and Cures*. IMF Staff Papers 45(4):559-594. International Monetary Fund.
26. The Hungarian Gallup Institute. (1999). *Basic Methodological Aspect of Corruption Measurement: Lessons Learned from the Literature and the Pilot Study (1999 December)*. URL: http://www.nobribes.org/Documents/corruption_hungary_wp_prelim.pdf
27. The Parliamentary Center of Canada. (2000). *Controlling Corruption: A Parliamentarian's Handbook 2nd edition*. Canada Parliamentary Center, World Bank Institute, Canadian International Development Agency.
28. Transparency International (2003). *Survey Sources for The Corruption Perspective Index.* Online Publications. URL: http://www.transparency.org/surveys/index.html
29. Xu, Chenggang. & Pistor, Katharina. (2002). *Law Enforcement under Incomplete Law: Theory and Evidence from Financial Market Regulation*. Discussion Paper No. TE/02/442. The Suntory Centre. London School of Economics and Political Science.

# **Appendix**

The basic values used in the simulations of the Artificial Corrupt Societies:

| | |
|---|---|
| Pay-off to the corruption ($\alpha$) | 20 |
| Pay-off to being honest ($\beta$) | 1 |
| Distribution of Honesty Index | Randomly Distributed (Gaussian Distribution) among agents |
| Number of Agents | 75 members of *P* and 75 members of *G* |
| Honesty Index | [0..1] |
| Length of jail period | 5 rounds |
| Social Network | 5 agents |
| Memory | 3 |
| Probability to be caught | 0.2 |
| Iterations | 200 rounds |

Some of the basic values are changed to see the effect to the evolutionary stable conditions in the whole artificial social system.